\documentclass[12pt]{article}

\pdfoutput=1

\usepackage{cite}

\usepackage{amssymb,latexsym}

\usepackage{epstopdf}

\usepackage{graphics,epsfig}

\usepackage{caption}

\usepackage{amsmath,amsfonts}
\usepackage{amssymb}
\usepackage{float}
\usepackage{amssymb,latexsym}

\usepackage{appendix}

\usepackage{graphicx}
\DeclareGraphicsExtensions{.pdf,.png,.jpg}

\usepackage{color}

\usepackage{mathrsfs}

\DeclareMathAlphabet{\mathpzc}{OT1}{pzc}{m}{it}

\let\a=\alpha \let\b=\beta \let\g=\gamma \let\d=\delta \let\e=\epsilon
  \let\th=\theta  \let\k=\kappa
\let\l=\lambda \let\m=\mu \let\n=\nu \let\x=\xi \let\p=\pi 
\let\s=\sigma   \let\f=\phi  
 
      \let\G=\Gamma \let\D=\Delta \let\Th=\Theta \let\L=\Lambda
\let\X=\Xi  \let\S=\Sigma  \let\Y=\Psi
 
\let\la=\label  
  
\def\nn{\nonumber} \def\bd{\begin{document}} \def\ed{\end{document}}
\def\ds{\documentstyle} \let\fr=\frac \let\bl=\bigl \let\br=\bigr
\let\Br=\Bigr \let\Bl=\Bigl
\let\bm=\bibitem
\let\na=\nabla
\def\tU{{\widetilde U}}
\let\pa=\partial \let\ov=\overline
\def\ie{{\it i.e.\ }}
\newcommand{\be}{\begin{equation}}
\newcommand{\ee}{\end{equation}}
\def\ba{\begin{array}}
\def\ea{\end{array}}
\def\ft#1#2{{\textstyle{{\scriptstyle #1}\over {\scriptstyle #2}}}}
\def\fft#1#2{{#1 \over #2}}
\def\F#1#2{{ F_{#1}^{(#2)} }}
\def\cF#1#2{{ {\cal F}_{#1}^{(#2)} }}

\def\R{{\bf R}}
\def\sst#1{{\scriptscriptstyle #1}}
\def\oneone{\rlap 1\mkern4mu{\rm l}}
\def\e7{E_{7(+7)}}
\def\td{\tilde}
\def\wtd{\widetilde}
\def\im{{\rm i}}
\def\bog{Bogomol'nyi\ }
\newcommand{\ho}[1]{$\, ^{#1}$}
\newcommand{\hoch}[1]{$\, ^{#1}$}
\newcommand{\bea}{\begin{eqnarray}}
\newcommand{\eea}{\end{eqnarray}}
\newcommand{\ra}{\rightarrow}
\newcommand{\lra}{\longrightarrow}
\newcommand{\Lra}{\Leftrightarrow}
\newcommand{\ap}{\alpha^\prime}
\newcommand{\bp}{\tilde \beta^\prime}
\newcommand{\cB}{{\cal B}}
\newcommand{\cO}{{\cal O}}
\newcommand{\vecx}{\vec{x}}
\newcommand{\vecy}{\vec{y}}
\newcommand{\vecp}{\vec{p}}
\newcommand{\vecq}{\vec{q}}
\newcommand{\tr}{{\rm tr} }
\newcommand{\Tr}{{\rm Tr} }
\newcommand{\NP}{Nucl. Phys. }

\newcommand{\cL}{{\cal L}}
\newcommand{\cA}{{\cal A}}
\newcommand{\cT}{{\cal T}}
\newcommand{\cD}{{\cal D}}
\newcommand{\cH}{{\cal H}}
\newcommand{\cR}{{\cal R}}

\def\sst#1{{\scriptscriptstyle #1}}
\def\0{{\sst{(0)}}}
\def\1{{\sst{(1)}}}
\def\2{{\sst{(2)}}}
\def\3{{\sst{(3)}}}
\def\4{{\sst{(4)}}}
\def\5{{\sst{(5)}}}
\def\6{{\sst{(6)}}}
\def\7{{\sst{(7)}}}
\def\8{{\sst{(8)}}}
\def\9{{\sst{(9)}}}
\def\p{{\sst{(p)}}}
\def\q{{\sst{(q)}}}
\def\ve{\varepsilon}
\def\vf{\varphi}
\def\F{\Phi}
\def\wg{\wedge}

\def\thb{\bar{\theta}}
\def\Thb{\bar{\Theta}}
\def\barp{\bar{p}}
\def\barq{\bar{q}}
\def\barc{\bar{c}}
\def\bard{\bar{d}}
\def\e{\epsilon}

\def \bi{\bibitem}
\def \la {\label}

\def \l {\lambda}
\def\foot{\footnote}
\def \tl  {{\tilde \l}}
\def \sql {{\sqrt \l}}
\def \adss {$AdS_5 \times S^5$\ }
\newcommand{\rf}[1]{(\ref{#1})}
\def \ov {\over}

\def\th{\theta}
\def\Th{\Theta}
\def\vth{\vartheta}
\def\btheta{{\bar\theta}}
\def\ttheta{{{\tilde\theta}}}
\def\bttheta{{{\bar\ttheta}}}
\def\vth{\vartheta}

\def\ra{\rightarrow}
\def\N{\nabla}
\def\F{{\cal F}}
\def\uM{\underline{M}}
\def\uA{\underline{A}}
\def\uN{\underline{N}}
\def\uP{\underline{P}}
\def\ua{\underline{a}}
\def\ub{\underline{b}}
\def\uc{\underline{c}}
\def\ud{\underline{d}}
\def\ue{\underline{e}}
\def\uf{\underline{f}}
\def\ui{\underline{i}}
\def\uj{\underline{j}}
\def\uk{\underline{k}}
\def\ul{\underline{l}}
\def\ual{\underline{\alpha}}
\def\ube{\underline{\beta}}
\def\um{\underline{m}}
\def\un{\underline{n}}
\def\up{\underline{p}}
\def\uq{\underline{q}}
\def\ur{\underline{r}}
\def\us{\underline{s}}
\def\umu{\underline{\mu}}
\def\unu{\underline{\nu}}
\def\ula{\underline{\l}}
\def\uka{\underline{\k}}
\def\usi{\underline{\s}}
\def\urh{\underline{\r}}
\def\cc{\circ}
\def\eqv{\equiv}

\def\ni{\noindent}

\def\Ep{E^{{}^{(+)}}}
\def\Em{E^{{}^{(-)}}}

\def\Mp{M^{{}^{(+)}}}
\def\Mm{M^{{}^{(-)}}}

\def \ha{{1\ov 2}}

\def\r{\rho}

\def\Y{{\rm Y}}
\def\X{{\rm X}}
\def\tY{\tilde{\rm Y}}
\def\tX{\tilde{\rm X}}
\def\dY{\dot{\rm Y}}
\def\dX{\dot{\rm X}}

\def \J {\mathcal{J}}
\def \del {\partial}

\def\dF{\dot{F}}
\def\dG{\dot{G}}
\def\df{\dot{f}}
\def \E {{\cal E}}
\def \S {{\cal S}}
\def \J {{\cal J}}

\def\ms{\mathcal{S}}
\def\mj{\mathcal{J}}
\def\soj{\fr{\ms}{\mj}}
\def \R {{\bf R}}
\def \om {\omega}
\def \bE {\bar E}
\def \x {{\cal X}}

\def \bi{\bibitem}
\def \la {\label}

\def \l {\lambda}
\def\foot{\footnote}
\def \tl  {{\tilde \l}}
\def \sql {{\sqrt \l}}
\def \adss {$AdS_5 \times S^5$\ }
\def \ov {\over}

\def \varpi {{\rm w}}

\def\thb{\bar{\theta}}
\def\Thb{\bar{\Theta}}
\def\mb{\bar{\m}}
\def\ab{\bar{\a}}
\def\zb{\bar{z}}
\def\psib{\bar{\psi}}
\def\barp{\bar{p}}
\def\barq{\bar{q}}
\def\barc{\bar{c}}
\def\bard{\bar{d}}
\def\e{\epsilon}
\def\wb{\bar{w}}
\def\lb{\bar{\l}}
\def\Jb{\bar{J}}
\def\Nb{\bar{N}}
\def\Zb{\bar{Z}}
\def\pab{\bar{\pa}}

\def\Cb{\bar{C}}

\def\At{\tilde{A}}
\def\Bt{\tilde{B}}
\def\Ct{\tilde{C}}
\def\Dt{\tilde{D}}
\def\Et{\tilde{E}}
\def\Ft{\tilde{F}}
\def\Gt{\tilde{G}}
\def\Ht{\tilde{H}}
\def\Kt{\tilde{K}}
\def\Mt{\tilde{M}}
\def\Nt{\tilde{N}}
\def\Rt{\tilde{R}}
\def\at{\tilde{a}}
\def\bt{\tilde{b}}
\def\ct{\tilde{c}}
\def\dt{\tilde{d}}
\def\et{\tilde{e}}
\def\ft{\tilde{f}}
\def\htil{\tilde{h}}
\def\gt{\tilde{g}}
\def\nt{\tilde{n}}
\def\mut{\tilde{\mu}}
\def\nut{\tilde{\nu}}
\def\pht{\tilde{\f}}
\def\vft{\tilde{\vf}}

\def\rht{\tilde{\rho}}

\def\asth{\hat{*}}
\def\phh{\hat{\phi}}

\def\bA{{\bf A}}

\def\kbf{\mathbf{k}}
\def\ybf{\mathbf{y}}
\def\xbf{\mathbf{x}}
\def\Lbf{\mathbf{L}}
\def\rbf{\mathbf{r}}

\def\ola{\overleftarrow}
\def\ora{\overrightarrow}
\def\alt{\tilde{\a}}

\def\eh{\hat{e}}
\def\eph{\hat{\e}}
\def\ph{\hat{p}}
\def\alh{\hat{\a}}
\def\beh{\hat{\b}}
\def\gah{\hat{\g}}
\def\Fh{\hat{F}}
\def\muh{\hat{\m}}
\def\nuh{\hat{\n}}
\def\thh{\hat{\th}}
\def\rhh{\hat{\r}}
\def\dh{\hat{d}}
\def\ih{\hat{i}}
\def\jh{\hat{j}}
\def\hh{\hat{h}}
\def\nh{\hat{n}}
\def\gh{\hat{g}}
\def\kh{\hat{k}}
\def\deh{\hat{\d}}
\def\wh{\hat{w}}
\def\lah{\hat{\l}}
\def\Ah{\hat{A}}
\def\Kh{\hat{K}}
\def\Nh{\hat{N}}
\def\Rh{\hat{R}}
\def\Ch{\hat{C}}
\def\Omh{\hat{\Omega}}

\def\xh{\hat{x}}

\def\ps{\rlap{\, /}\;\,p }
\def\ks{\rlap{\, /}\;\,k }

\def\gym{g_{YM}}

\def\adot{\dot{a}}
\def\bdot{\dot{b}}
\def\bpa{\bar{\pa}}

\def\pr{\prime}
\def\ssk{\medskip}
\def\clb{\color{blue}}
\def\clr{\color{red}}
\def\clg{\color{green}}

\begin{document}

\overfullrule=0pt
\parskip=2pt
\parindent=12pt
\headheight=0in \headsep=0in \topmargin=0in
\oddsidemargin=0in

\vspace{ -3cm}
\thispagestyle{empty}

 \vspace{0.1cm}

\setcounter{equation}{0}
\setcounter{footnote}{0}
\setcounter{section}{0}

\begin{center}

{\Large\bf  Holographic quantization of gravity in a black hole background}

\vskip 0.8cm

 \vspace{.5cm}

\vspace{0.5cm}
I. Y. Park
\\

\vspace{0.3cm}



\vspace{0.3cm}
{\it Department of Applied Mathematics,
Philander Smith College 
                               \\
Little Rock, AR 72223, USA \\
inyongpark05@gmail.com
}

\end{center}

 \vspace{0.1cm}

\begin{abstract}

\ni It was recently observed in \cite{Park:2014tia} that the holographic nature of gravity may hold a key to quantization of gravity. The so-called ``holographic quantization" has been carried out in \cite{Park:2014noa,Park:2015ota} for Einstein gravity in a flat background. Generalizing the procedure to a curved background is the main goal of the present work. In particular, we consider the Einstein action expanded around a Schwarzschild background.

\end{abstract}
\newpage
\section{Introduction}

When a bulk spacetime admits a foliation, its dynamics may be described through 
the collective dynamics of the leaves, the co-dimension one hypersurfaces.  
Although the total dynamics of the leaves and the direction transverse to the leaves is a priori expected to have the bulk degrees of freedom, it has turned out that it has only the surface degrees of freedom, the holographic property of gravity \cite{York:1972sj}\cite{'tHooft:1993gx}, which in turn has its origin in the large amount of gauge symmetry.
By employing the ADM formulation \cite{Arnowitt:1960es}, it has recently been shown \cite{Park:2014tia,Park:2014qoa} that the holographic nature of gravity may be a key to the long-standing problem of quantization of gravity \cite{DeWitt:1967ub,'tHooft:1974bx,Deser:1974cz,Goroff:1985th,Anselmi:2003xb,Stelle:1976gc,
Antoniadis:1986tu,Weinberg3,Reuter:1996cp}. The procedure dubbed ``holographic quantization" has been carried out for Einstein gravity in a flat background \cite{Park:2014tia,Park:2014noa,Park:2015ota}. Generalizing the procedure to a curved background is the main goal of this paper.\footnote{The equivalence between the usual formulation and ADM formulation of general relativity was questioned in \cite{Kiriushcheva:2008sf}. The task undertaken in this work is the curved space analogues of \cite{Park:2014tia,Park:2015ota}: quantization of gravity associated with the physical states dictated by the ADM formulation with the synchronous type gauge.}

In comparison with the existing approaches to quantization, our goal is more modest in 
two aspects.
A YM theory has the very nice property that it is renormalizable even if the gauge modes are placed on the external lines of the Feynman diagrams. In other words, the theory is renormalizable completely off-shell. 
As far as we can see, a gravity theory does not seem to share the same luxury, at least not in any obvious way.
It has been proposed in \cite{Park:2014tia} that a slightly modest goal of renormalizing the physical states can be achieved. Secondly, instead of the goal of finding a background-independent quantization we aim at a less ambitious goal of a background-specific quantization that has
moderate background dependence. For example, the gauge choices available should depend on the background under consideration as we will analyze later.

A curved spacetime requires, not surprisingly, more care compared with the flat background analyses\footnote{Since renormalizability should be a local property, the curved space case is expected to work as well. However, the details of the renormalization procedure would be needed in applications to various black hole physics.} in \cite{Park:2014noa} and \cite{Park:2015ota}: one should be more careful about gauge-fixings, the reduction procedure is more subtle and the computation of the propagator more challenging. (Fortunately, only the tensor structures of the propagators will be needed for calculating the counter-terms.) For example, the flat space gauge-fixing $K=0$ where $K$ denotes the trace of the second fundamental form $K_{mn}$ should be modified. Nevertheless, the procedure can be carried out and we illustrate the quantization 
with a Schwarzschild black hole background.

As a matter of fact, the gauge-fixing of $K$ turns out to bear an interesting origin.
The flat space fixing $K=0$ is generalized to $K=K_\0$ - the condition that $K$ be non-dynamical - where $K_\0$ denotes the classical value. Recall that the condition $K=0$ appeared through one of the nonlinear de Donder gauge equations and played a crucial role in the flat space quantization. More generally, $K=K_\0$ arises from gauging away the trace part of the fluctuation metric: the condition $K=K_\0$ is a constraint that arises from the gauge-fixing of the conformal-transformation part of the diffeomorphism, as will be analyzed in section \ref{Kgauge}.
The constraint will again play a critical role analogous to that in the flat case.

The $K=K_\0$ condition eventually leads to the reduction of the second fundamental form $K_{mn}$ in a manner that generalizes the flat case. (Previous works in which the relevance of a hypersurface as true degrees of freedom was noticed include \cite{York:1972sj,Moncrief:1989dx,Fischer:1996qg,Gay-Balmaz:2014ena}.) Once the reduction is established, the renormalization can be carried out similarly to the flat case. 
One of the complications is that the expression of the 4D Riemann tensor in terms of the 3D quantities contains more terms than in the flat case. We revisit the renormalization of the graviton one-loop two-point amplitude and also comment on other amplitudes.

\vspace{.4in}
\ni The rest of the paper is organized as follows.\\

\ni In section 2, we review the flat space quantization and set the stage for the curved space analysis. Some of the steps in the flat space quantization become generalized and come to carry clearer meanings in the systematic curved space analysis in section \ref{Kgauge}. In particular,  we conduct an in-depth analysis of the $K=0$ gauge. 
We first slightly refine our gauge-fixing strategy because of the new element, the gauge-fixing of the trace piece of the fluctuation metric. We fix the shift vector as before; in consequence the already-nondynamical $n$-field (whose non-dynamism is well known through the Hamiltonian analysis) becomes further constrained (see \rf{pmnz} below). 
We then gauge-fix the conformal part of the diffeomorphism.

After reviewing the literature that points out the unphysical nature of the trace of the fluctuation metric, we gauge away the trace. The field equation of the trace piece must be introduced as a constraint which is nothing but the $K$-constraint. 
Then we show that the $K=K_\0$ constraint leads to the reduction of the second fundamental form $K_{mn}$. The mathematical duality between the Riemannian foliation and totally geodesic foliation will play a crucial role. The interplay between the Lagrangian and Hamiltonian formulations will be important.
 The reduction of the effective action - the curved space analogue of the analysis carried out in \cite{Park:2015ota} - is outlined in section 4. As in the flat case, the theory is shown to be renormalizable by a metric field redefinition at the end. We end with a summary and future directions.
Appendix A has our conventions and useful identifies. We show in Appendix B that a certain part of the analysis readily applies, with a slight modification, to the AdS and dS black hole cases whose full analyses are left for the future. In Appendix C, the reduction of $K_{mn}$ is further elucidated by taking an example of a scalar theory.

\section{Review of flat case}

It will be useful to recall the flat space quantization to set the stage for the curved case analysis. The flat case has many of the ingredients of the curved space but is simpler. Consider the Einstein-Hilbert action
\bea
S=\int d^4 x \sqrt{-g}\;R \la{EH}
\eea
Let us introduce the fluctuation metric $\f_{\m\n}$:
\bea
g_{\m\n}\equiv \eta_{\m\n}+\f_{\m\n}
\eea
One can expand the action around a flat metric; to the second order of the fluctuation fields it is given by (the action has been rescaled by an overall factor 2) 
\bea
 S &=& \sqrt{-g}\Big( -\fr12{\N}_\g \f^{\a\b}{\N}^\g \f_{\a\b}+\fr12 {\N}_\g \f^{\a}_\a {\N}^\g \f^{\b}_\b+ \nabla_\k \f^{\k\m}\nabla^\l \f_{\l\m}-\nabla^\l \f_{\l\m} \nabla^\m \f_\a^\a \la{gravcubcov}
\\
&&+\f_{\a\b}\f_{\g\d}R^{\a\g\b\d}-\f_{\a\b}\f^{\b}{}_\g R^{\k\a\g}{}_{\k}
-\f^{\a}{}_{\a}\f_{\b\g}R^{\b\g}-\fr12 \f^{\a\b}\f_{\a\b}R
+\fr14  \f^{\a}_\a  \f^{\b}_\b R +\cdots\Big) \nn
\eea
where the connection and the fields $R, R_{\m\n}, R_{\m\n\r\s}$ are functions of the background metric and of course vanish for a flat background. However, the form of the action above is useful for the refined application of the background field method as explained in \cite{Park:2015ota}.
The gauge-fixing term (again rescaled by a factor 2) is given by
\bea
\cL_{g.f.}=-\Big(\f_\m-\fr12 \nabla_\m \f\Big)^2;
\eea 
the corresponding ghost term should be added to the action above. 
Renormalizability can be established once one considers only the physical external states in the Feynman diagrams. Let us now take a short excursion to the Hamiltonian formalism for determination of the physical states. (The interplay between the Lagrangian and Hamiltonian will be even more important in the curved space analysis.)

A globally hyperbolic spacetime admits separation of one of the spatial coordinates $x^3$ from the rest 
\bea
x^\m\equiv (y^m,x^3)
\eea
where $\m=0,..,3$ and $m=0,1,2$ 
and parameterization of the metric (see, e.g., \cite{Wald,Poisson,Blau,Embacher:1995xp} for reviews)
\bea
g_{\m\n}=\left(
\begin{array}{cc}
{ h_{mn}} & N_{ m} \\
&\\
N_{ n} &  n^2+h^{mn}N_{m} N_{ n} 
\end{array}
\right)\quad,\quad
g^{\m\n}=\left(
\begin{array}{cc}
h^{mn}+\fr1{n^2}N^m N^n & -\fr1{n^2}N^{ m} \\
&\\
-\fr1{n^2}N^{ n} &  \fr1{n^2} 
\end{array}
\right)
\eea
We consider a Schwarzschild black hole later. Only the outside of the black hole is globally hyperbolic. However, it should not matter since the renormalizability should be a local property.
The coordinate $x^3$ will be taken as one of the spatial coordinates for the flat case at hand and as the radial coordinate $r$ for the black hole case later. Even though these coordinates are not the genuine time coordinates, we still call the spacetimes ``globally hyperbolic."
In the ADM variables, the action takes (the total derivative terms will not be kept track of for now) 
\bea
S=\int d^4 x\;n\sqrt{-h} \left(\cR+K^2-K_{mn}K^{mn}
\right)
\la{1p3act}
\eea
where $\cR$ denotes the 3D Ricci scalar 
\bea
\cR:\;\;\mbox{3D Ricci scalar}
\eea
(The 3D Ricci tensor and Riemann tensor will be denoted by $\cR_{mn}, \cR_{mnpq}$ respectively) and
\be
K_{mn}=\fr1{2n}\left(\mathscr{L}_{\pa_{x^3}} h_{mn}-{\nabla}_m N_{n}
         -{\nabla}_n N_{ m} \right),\qquad K\equiv h^{mn}K_{mn}.
\la{K4defqq}
\ee
where $\mathscr{L}_{\pa_{x^3}}$ denotes the Lie derivative along the vector field $\pa_{x^3}$ and $\N_m$ is the 3D covariant derivative constructed out of $h_{mn}$; $n$ and $N_{ m}$ denote the lapse function and shift vector, respectively. The ADM form of the action reveals that they are non-dynamical. One can define the canonical momentum and straightforwardly compute the Hamiltonian. The field equations of the lapse and shift can also be easily computed in the standard manner and are imposed as constraints. They can be expressed in terms of the Lagrangian variables:
\bea
\cR-K^2+K_{mn}K^{mn}=0  \la{ncon}
\eea
\bea
{\N}_m (K^{mn}-h^{mn} K)=0  \la{Ncon}
\eea
As shown in \cite{Park:2014tia} and \cite{Park:2014qoa}, the shift vector constraint \rf{Ncon}
implies
\bea
\pa_m n=0 \la{pmnz}
\eea
All of these (other than \rf{pmnz}) are standard; more details will be presented in the curved analysis for self-containedness of the paper. Let us come back to the Lagrangian description. What is important for now is the fact that the metric field equation combined with the constraints above leads to
\bea
\cR=0\quad,\quad -K^2+K_{mn}K^{mn}=0 \la{r3kk}
\eea
This has been shown in several different ways in the previous works \cite{Park:2014tia}\cite{Park:2015qxa}.

The nonlinear de Donder gauge $g^{\r\s}\G^\m_{\r\s}=0$ \cite{Smarr:1978dia}\cite{Blau} can be imposed; it reads, in the ADM fields,
\bea
&& (\pa_{x^3}-N^m \pa_m) n=n^2K  \nn\\
&& (\pa_{x^3}-N^n \pa_n)N^m=n^2(h^{mn}\pa_n \ln n-h^{pq}\G^m_{pq})
\la{ADMdd}
\eea
Several remarks are in order. As shown in \cite{Park:2014tia,Park:2014qoa}, the shift vector constraint \rf{Ncon} combined with the first equation of \rf{ADMdd} and the non-dynamism of $n$ implies $K=0$. (This `gauge' seems unnatural, however, in the black hole case, since the background does not satisfy the gauge. 
We will discuss a generalization of this gauge in the next section. 
It turns out that $K=0$ is actually a constraint associated with the unphysical trace piece of the metric.)

The $K=0$ gauge leads to the reduction of the second fundamental form $K_{mn}$.
To see this, let us first combine the $n$-field equation with trace of $h^{mn}$ field equation.
The $h^{mn}$ field equation is
\bea
&& n(\cR_{mn}-\fr12h_{mn}\cR)+\fr12 n h_{mn}K^2 +h_{mn}{\pa_{3}}K
  +\fr12  nh_{mn}K_{rs}K^{rs} \nn\\
&&-2 nK_{mp}K^p{}_n - nK_{mn}K
-h_{mp}h_{nq}{\pa_{3}}K^{pq}=0
\label{gmneomflat}
\eea
Multiplication of $h^{mn}$ yields
\bea
-\fr12 n \cR+\fr12nK^2-\fr12 n K_{rs}K^{rs}+3\pa_3 K-h_{pq}\pa_3 K^{pq}=0
\eea  
Combining this with the $n$-field constraint implies
\bea
&&3\pa_3 K-h_{pq}\pa_3 K^{pq}
=2\pa_3 K+2n K_{pq}K^{pq}=0
\eea 
Upon using the K-gauge $K=0$, one gets
\bea
K_{pq}K^{pq}=0
\eea
which, after Wick rotation, implies $K_{mn}=0$.

Before turning to the analysis of the curved case, let us now review the renormalization of the two-point graviton amplitude in Fig \ref{graviBFM}. (We refer to \cite{Park:2015ota} for more details.)
\begin{figure}[tbp]
\centering 
\includegraphics[width=1.1\textwidth,trim=50 640 -10 15,clip]{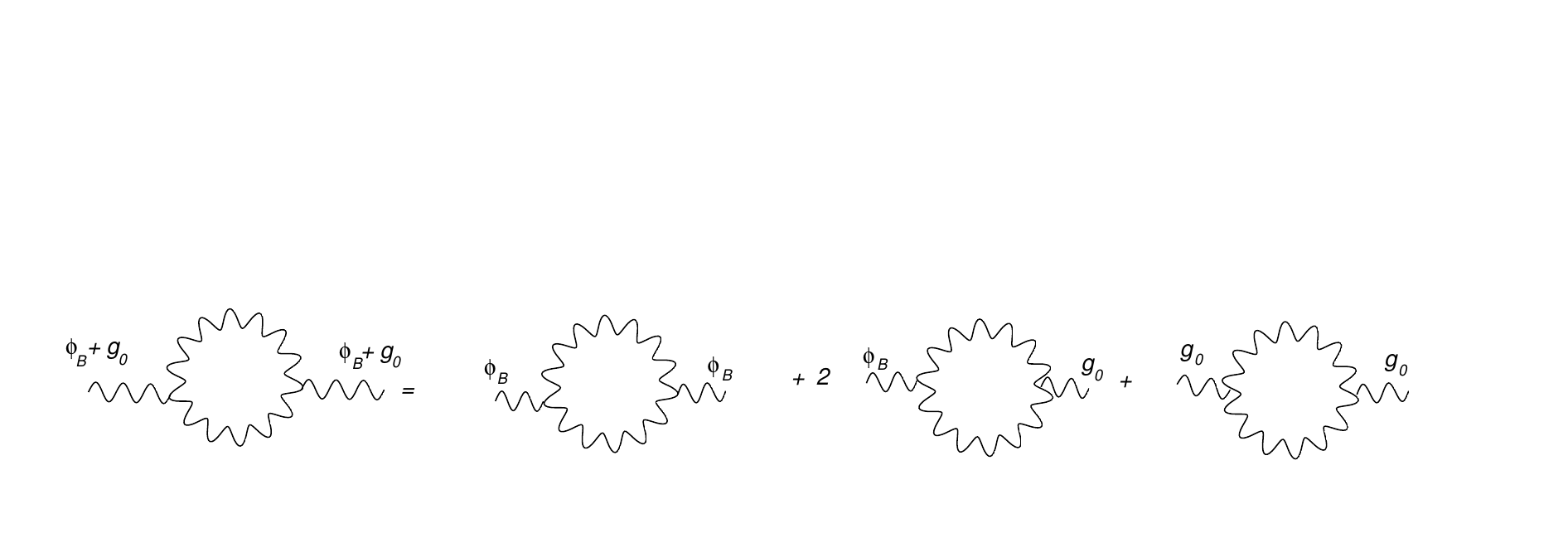}
\caption{BFM for graviton one-loop (spacetime indices suppressed)}
\la{graviBFM}
\end{figure}
The 4D graviton propagators is given by
\bea 
<\f_{\m\n}(x_1)\f_{\r\s}(x_2)>&=& P_{\m\n\r\s} \int \fr{d^4k}{(2\pi)^4}\fr{e^{ik\cdot (x_1-x_2)}}{i k^2} 
\eea
where, for the traceless propagator,
\bea
P_{\m\n\r\s}\equiv \fr12(\eta_{\m\r}\eta_{\n\s}+\eta_{\m\s}\eta_{\n\r}- \fr12\eta_{\m\n}\eta_{\r\s})
\eea
Let us shift the metric:  
\bea
g_{\m\n}\equiv  \f_{\m\n}+\tilde{g}_{{}_B\m\n}\quad \mbox{where}\quad \tilde{g}_{{}_B\m\n}\equiv \vf_{{}_B\m\n}+g_{0\m\n},\quad g_{0\m\n}=\eta_{\m\n};
\eea
The graviton sector correlator is given by
\bea
&&\hspace{-.7in}-\fr12<\Big(\f_{\a\b}\f_{\g\d}\Rt^{\a\g\b\d}-\f_{\a\b}\f^{\b}{}_\g \Rt^{\k\a\g}{}_{\k}
-\f^{\a}{}_{\a}\f_{\b\g}\Rt^{\b\g}-\fr12 \f^{\a\b}\f_{\a\b}\Rt
+\fr14  \f^{\a}_\a  \f^{\b}_\b \Rt \Big)^2> \nn\\
\hspace{.2in}&=&-\fr12<\Big(\f_{\a\b}\f_{\g\d}\Rt^{\a\g\b\d}-\f_{\a\b}\f^{\b}{}_\g \Rt^{\k\a\g}{}_{\k}
-\fr12 \f^{\a\b}\f_{\a\b}\Rt
\Big)^2> 
\eea
where the equality is due to our use of the traceless propagator.
The fields with tildes are functions of 
$\tilde{g}_{{}_B\m\n}\equiv \eta_{\m\n}+\vf_{{}_B\m\n}$.
After some algebra of integrating out $\f_{\m\n}$, one gets, for the counter-terms, (the tildes have been omitted)
\bea
\D \cL
&=&-\fr12\fr{\G(\e)}{(4\pi)^2}\Big[-3R_{\m\n}R^{\m\n}-R^2\Big]  \la{tgravi}
\eea
The analysis for the ghost sector is parallel (but it reveals the necessity of 
the gauge-fixing of the trace part \cite{Park:2015ota}).

\section{$K=K_\0$ constraint \la{Kgauge}}

For the curved space generalization of $K=0$ condition and other purposes, let us keep careful track of the boundary terms.
The variation of the Einstein-Hilbert action leads to the equation-of-motion terms and the boundary terms that can be put into the form of 
\bea
-2\int_\pa  K \la{Kbd}
\eea  
where $\pa$ denotes the boundary of the spacetime under consideration.
Conventionally the following ``counter boundary terms" are added in order to cancel the boundary term above and thereby enable one to impose the Dirichlet boundary conditions on the metric components (see, e.g., \cite{Poisson}):
\bea
2\int_\pa  K-2\int_\pa  K_\0
\eea
where $K_\0$ denotes the value of $K$ for the background metric.
 The addition of $\int_\pa  K_\0$ is to make the action finite. With these boundary terms added, the action takes
\bea
S_{\mbox{bulk+bd}}=\int d^4 x \sqrt{-g}\;R+2\int_\pa  K-2\int_\pa  K_\0
\eea
The boundary term will be carefully examined below in the Lagrangian analysis. The analysis unravels an interesting aspect of the $K=0$ gauge or its generalization, the $K=K_\0$ constraint.
The meaning of this constraint becomes even clearer in the subsequent ADM canonical Hamiltonian analysis.

\subsection{unphysical nature of trace of metric \la{upn}}

As stated previously, the constraint $K=K_\0$ arises from gauge-fixing of the trace piece of the fluctuation metric. The full implication of the gauge-fixing of the trace piece of the fluctuation metric
can be understood through the interplay of the Lagrangian and Hamiltonian analyses. It is easier to see the gauge symmetry associated with the trace piece in the Lagrangian formalism; this formalism reveals that one should introduce a constraint on $K$ at the boundary. (One may say that one sees only a hint for the constraint since the boundary term has been subtracted out; see more comments below.) The Hamiltonian analysis leads to the bulk version of the $K$-constraint.

Before we get to the Lagrangian and Hamiltonian analyses, let us recall the literature on the trace piece of the fluctuation metric \cite{Kuchar:1970mu}\cite{Gibbons:1978ac}\cite{Mazur:1989by}.
The salient point was that the diffeomorphism contains a special form of the conformal transformation which, therefore, should be removed by gauge-fixing.
This conformal transformation is associated with the trace part of the fluctuation metric.
To see this in detail, let us recast the diffeomorphism transformation with a parameter $\xi^\m$,
$\d g_{\m\n}=\nabla_\m \xi_\n+\nabla_\n \xi_\m$,
into the form
\bea
\d g_{\m\n}=\fr12 (\nabla_\k \xi^\k)g_{\m\n}+(Lg)_{\m\n} \la{diffeo}
\eea
where $(Lg)_{\m\n}$ denotes the traceless part of the Lie derivative.
\bea
(Lg)_{\m\n}\equiv \nabla_\m \xi_\n+\nabla_\n \xi_\m-\fr12 (\nabla_\k \xi^\k)g_{\m\n}
\eea
As one can see, the first term of \rf{diffeo} takes the form of a conformal transformation.
Since this particular conformal transformation is part of the gauge symmetry (i.e., the diffeomorphism), it must be fixed.

\subsubsection*{Lagrangian and Hamiltonian analyses}

Above we have seen that the trace piece of the metric is associated with the conformal part of the diffeomorphism. The trace piece also makes the path integral ill-defined as one can see as follows. 
We will shortly derive the background covariant expansion of the action to the quadratic order of the fluctuation, \rf{gravcubcov}. For now let us consider the flat case and the expansion
\bea
\sqrt{-g}\;\cL = \sqrt{-g}\Big( -\fr12{\N}_\g \f^{\a\b}{\N}^\g \f_{\a\b}+\fr14 {\N}_\g \f^{\a}_\a {\N}^\g \f^{\b}_\b\Big)  \la{gravcubcovf}
\eea
and not insist on the advantage of covariance brought by the redefined application of the background field method. As noticed in the literature, the presence of the second term 
makes the path integral ill-defined and divergent.\footnote{There has been progress in dealing with the `ghost' modes (the modes with wrong-sign kinetic terms) in the gravity with infinite number of higher derivative terms \cite{Biswas:2011ar}. } 

The path integral for the curved case becomes ill-defined basically for the same reason.
Let us define (see Appendix A for our conventions)
\bea
g_{\m\n}=g_{0\m\n}+\f_{\m\n}
\eea
where $g_{0\m\n}$ denotes the Schwarzschild metric,
\bea
ds^2= -f(r)dt^2+\fr1{f(r)}dr^2+r^2{(d\th^2+\sin^2\th d\varphi^2)}
\eea
with
\bea
f(r)\equiv {1-2M/r}
\eea
One can prevent $\f$ from causing such pathology by gauge-fixing. At the linear level the gauge-fixing is 
\bea
g_0^{\m\n}\f_{\m\n}=0  \la{lgf}
\eea 
After completing the linear level analysis, we will take up the task of nonlinear generalization of this gauge-fixing. We will now show that the condition $K=K_\0$ is nothing but the corresponding constraint.

We start with the usual covariant (i.e., non-ADM) Lagrangian formalism because it is this formalism that easily reveals the non-dynamism of the trace piece. The constraint $K=K_\0$ arises from the boundary term that one subtracts out in order to impose the Dirichlet boundary conditions. Therefore one may say that one sees only a hint for the constraint $K=K_\0$.\footnote{If one does not subtract out the boundary  term, it is not just a hint: it is a fact. }

For a general background, the expansion to the quadratic order is given by\footnote{Compared with \cite{Goroff:1985th}, the ${\f^{\a}{}_{\a}\f_{\b\g}}R^{\b\g}$-term appears with an opposite sign. The convention of \cite{Goroff:1985th} is such that
$R_{\m\n}\equiv R^\k{}_{\m \n\k}$, which differs from ours by an overall minus sign.} 
\bea
\hspace{.7in} &&\sqrt{-g}\;\cL = \sqrt{-g}\Big( -\fr12{\N}_\g \f^{\a\b}{\N}^\g \f_{\a\b}+\fr14 {\N}_\g \f^{\a}_\a {\N}^\g \f^{\b}_\b  \la{gravcubcov}
\\
&&\hspace{-1in}+\f_{\a\b}\f_{\g\d}R^{\a\g\b\d}-\f_{\a\b}\f^{\b}{}_\g R^{\k\a\g}{}_{\k}
 -\f^{\a}{}_{\a}\f_{\b\g}R^{\b\g}-\fr12 \f^{\a\b}\f_{\a\b}R
+\fr14  \f^{\a}_\a  \f^{\b}_\b R +\cdots\Big) \nn
\eea
Because we will make an intensive use of the functional derivative method in deriving the $K$-constraint, we illustrate derivation of some of the terms in \rf{gravcubcov} by the functional derivative method as a warm-up exercise. The functional Taylor expansion of the action \rf{EH} around a background $g_{0\m\n}$ is given by 
\bea
S(g_{0\m\n}\!\!+\!\f_{\m\n})\!
\!&=&\!S(g_{0\m\n})+\fr12\int\int \f_{\a\b}\f_{\m\n}\left.\fr{\d }{\d g_{\a\b}}\fr{\d S}{\d g_{\m\n}}\right|_{g_{\r\s}=g_{0\r\s}}+\cdots \nn\\ \la{fte}
\eea
where the linear term has been omitted since $g_{0\m\n}$ is taken to be a solution of the field equation.
Among the quadratic terms that appear in the expansion,
\bea
\fr12\int\int \f_{\a\b}\f_{\m\n}\fr{\d }{\d g_{\a\b}}\fr{\d S}{\d g_{\m\n}}
=\fr12\int \f_{\a\b}\f_{\m\n}\fr{\d }{\d g_{\a\b}}\Big[-\sqrt{-g}\;(R^{\m\n}-\fr12 g^{\m\n}R)\Big], \nn\\  \la{fe}
\eea
the most complicated contribution comes from
\bea
&&\int \f_{\a\b}\f_{\m\n}\sqrt{-g}\;\fr{\d }{\d g_{\a\b}}R^{\m\n} \nn\\
&=&\int \f_{\a\b}\f_{\m\n}\sqrt{-g}\;\Big[R^{\m\m'\n\n'}\fr{\d }{\d g_{\a\b}}g_{\m'\n'}
+g_{\m'\n'}\fr{\d }{\d g_{\a\b}}R^{\m\m'\n\n'}\Big]
\eea
The first term contributes to the first Riemann tensor term in \rf{gravcubcov}; the second term can be further expanded:
\bea 
\int \f_{\a\b}\f_{\m\n}\sqrt{-g}\;g_{\m'\n'}\fr{\d }{\d g_{\a\b}}R^{\m\m'\n\n'}
\eea
\[
\!\!\!=\!\!\!\int \f_{\a\b}\f_{\m\n}\sqrt{-g}\;g_{\m'\n'}\Big[R^{\m}{}_{\k_1\k_2\k_3}\fr{\d }{\d g_{\a\b}}g^{\k_1\m'}g^{\k_2\n}g^{\k_3\n'}+g^{\k_1\m'}g^{\k_2\n}g^{\k_3\n'}\fr{\d }{\d g_{\a\b}}R^{\m}{}_{\k_1\k_2\k_3}\Big]
\]
The computation of the first term is straightforward and it contributes to the Riemann tensor terms. The second term requires more care; it can be computed by using \rf{rtv} and it contributes to the first line of \rf{gravcubcov}. (To get the first line of \rf{gravcubcov}, the contribution from $\fr{\d R}{\d g_{\a\b}}$ in \rf{fe} and the gauge-fixing term should be taken into account as well.)

Let us get to one of the main tasks of this subsection: derivation of the $K$-constraint.
The field equation of the trace piece $\f$ should be imposed as a constraint once it is gauge-fixed:
\bea
\fr{\d S}{\d \f}=0 \la{fconstr}
\eea
Note that
\bea
\fr{\d S}{\d \f}=\Big[\fr{\d \f}{\d \f_{\m\n}(x)}\Big]^{-1}\fr{\d S}{\d \f_{\m\n}(x)}
=\Big[\fr{\d \f}{\d \f_{\m\n}(x)}\Big]^{-1}\fr{\d S}{\d g_{\m\n}(x)}
 =g_{0\m\n} \left[\fr{\d S}{\d g_{\m\n}(x)}\right]  \la{dsdf}
\eea
where $\fr{\d S}{\d \f}$ is taken with the traceless part of the metric fixed,
and
\bea
\d S=\int d^4x' R_{\a\b}{\d (\sqrt{-g}g^{\a\b})}
+ \int d^4x' \sqrt{-g}\,g^{\a\b}{\d R_{\a\b}}
\eea
The first term in the right-hand side yields the Einstein equation and the second term is the boundary term. We keep careful track of the boundary terms that come from the second term.  
Let us examine the boundary term in detail:
\bea
  \int d^3y \sqrt{-h}\;n^\k\Big[g^{\r\s}\nabla_\s \d g_{\r\k} -g^{\r\s}\nabla_\k \d g_{\r\s}\Big]
\eea 
The first term vanishes due to our gauge-fixing $n=n_0, N_m=0$:

\bea
\fr{\d S}{\d g_{\m\n}(x)}&\Rightarrow & \int d^4x' \sqrt{-g}\,g^{\a\b}\fr{\d R_{\a\b}}{\d g_{\m\n}(x)}
\nn\\
&=&\int d^4x' \sqrt{-g}\,\Big[-g^{\m\n} \nabla^\r\nabla_\r \d^\4(x'-x)\Big]
\nn\\
&=&- \int d^3y' \sqrt{-h}\, g^{\m\n} n^{\r} \nabla_\r\d^\4(x'-x)
\eea 
By partially integrating one gets
\bea
&=&\int d^3y' \sqrt{-h}\, g^{\m\n} \d^\4(y'-x)\nabla_{\r}n^{\r} 
\eea 
The upshot of the analysis so far is that
\bea
\fr{\d S}{\d \f}
              & \Rightarrow & \;(g_{0\m\n}g^{\m\n})\d(x_0^3-x^3)\,\nabla_{\r}n^{\r} 
\eea
where $x_0^3$ denotes the boundary value of $y^3$. 
Note that $\sqrt{-h}$ is no longer present: it has been integrated with $\d^{3D}(y'-x)$.
Since the Dirichlet boundary conditions have been imposed, $\f_{\m\n}$ vanishes at the boundary and $g^{\m\n}$ can be replaced by $g_0^{\m\n}$:
\bea
& \Rightarrow & \;(g_{0\m\n}g_0^{\m\n})\d(x_0^3-x^3)\,\nabla_{\r}n^{\r} 
\eea
The factor $\nabla_{\r}n^{\r}$ is another expression for $K$,
\bea
\nabla_{\r}n^{\r}=K 
\eea
Since the term $\int K_\0$, which should really mean
\bea
\Big[\int_\pa K\Big]|_{g_{\m\n}=g_{0\m\n}},
\eea 
is added as well in order to make the action finite, we impose
\bea
\d(x_0^3-x^3)\,K-\d(x_0^3-x^3)\,K_0=0
\eea
where the quantities $K,K_0$ take their boundary values. From this the sought-for condition $K-K_\0$ results:
\bea
K=K_\0 \mbox{ at boundary}  \la{KKzb}
\eea
Our next task is to derive a bulk version of \rf{KKzb}.

\vspace{.3in}

The bulk version of the $K$-constraint can be easily seen through the ADM Hamiltonian analysis. In the ADM variables, the action takes\footnote{Strictly speaking, the total derivative terms should be included:
\bea
R=\cR+K^2-K^{ab}K_{ab}+2\nabla_\a(n^\b \nabla_\b n^\a-n^\a \nabla_\b n^\b)
\eea
The total derivative terms should be distinguished from the boundary terms discussed previously:  they  (i.e., the total derivative terms) are present before taking any variation.
 The standard Hamiltonian analysis including the total derivative terms can be found, e.g., in \cite{Poisson}. The total derivative terms will be carefully analyzed for the present purpose in the nonlinear-level Lagrangian analysis that soon follows. 
} 
\bea
S=\int d^4 x\;n\sqrt{-h} \left(\cR+K^2-K_{mn}K^{mn}
\right)
\la{1p3actq}
\eea
Define the 3D metric fluctuation $q_{mn}$ through 
\bea
h_{mn}\equiv h_{0mn}+q_{mn} 
\eea
First we carry out the linear level analysis. Once we become more familiar with the setup through the linear exercise, we carry out the full analysis. (The crux of the nonlinear analysis is that there exists a nonlinear gauge whose linear part is the ADM analogue of \rf{lgf}.)

Let us expand 
\bea
K_{mn}=K_{\0 mn}+k_{ mn}
\eea
where $k_{mn}$ is the fluctuation part and adopting $N_m=0$ gauge it is given by
\bea
k_{mn}=
\fr1{2n}\mathscr{L}_{\pa_{x^3}} q_{mn}
\la{K4defqqlin}
\eea
The canonical momentum conjugate to $q_{mn}$ is given by
\bea
\pi^{mn}=\fr{\pa (\sqrt{-g}\,\cL)}{\pa \dot{q}_{mn}}=\sqrt{-h}({ -}K^{mn} { +}K h^{mn})
\la{canmtm}
\eea
where the dot is a shorthand notation for the Lie derivative $\mathscr{L}_{\pa_{x^3}}$ acting on $q_{mn}$:
\bea
\dot{q}_{mn}\equiv \mathscr{L}_{\pa_{x^3}q_{mn}}
\eea
It follows
\bea
\pi =2 \sqrt{-h}\; K\quad,\quad K^{mn}=\fr{1}{\sqrt{-h}}(-\pi^{mn}+\fr12 \pi h^{mn})
\eea
One can go to the Hamiltonian formulation by the usual Legendre transformation; the bulk part of the ``Hamiltonian of $x^3$-evolution" is
\bea
\cH\equiv  \pi^{mn}\dot{q}_{mn}- \sqrt{-g}\,\cL=\pi^{mn}(\dot{h}_{mn})- \sqrt{-g}\,\cL-\pi^{mn}\dot{h}_{0mn}
\eea
\[
=nh^{-1/2}( -\pi^{mn}\pi_{mn}  + \fr12 \pi^2)
-n h^{1/2}\cR-2N_m h^{1/2}\N_n(h^{-1/2}\pi^{mn}) -\pi^{mn}\dot{h}_{0mn}
\]
For any field $U$ including $n,N_m,\g_{mn}$, the Hamiltonian equation of motion can be written as 
\bea
 \mathscr{L}_{\pa_{x^3}}U=[U,H]_{PB}
 \la{peom}
\eea
where `PB' stands for the Poisson bracket.
The fact that $K=K_\0$ can be taken for the entire range of $x^3$ follows from the 
commutator of
\bea
[q_{mn},H]_{PB}=\mathscr{L}_{\pa_{x^3}}q_{mn}
\eea
Multiply $h_0^{mn}$ on both sides
\bea
h_0^{mn}[q_{mn},H]_{PB}= h_0^{mn}\dot{q}_{mn} \la{hzqd}
\eea
The gauge $g_0^{\m\n}\f_{\m\n}=0$ in the 4D covariant formulation translates into 
\bea
h_0^{mn}q_{mn}=0
\eea
since, after gauge-fixing, the lapse function takes $n=n_0(r)$ without any fluctuation part;
by setting $h_0^{mn}q_{mn}=0$, the left-hand side of \rf{hzqd} vanishes identically. Therefore one gets
\bea
h_0^{mn}\dot{q}_{mn}= 0 \la{qdot}
\eea
On account of the gauge $N_m=0$, the desired result follows from \rf{qdot}:
\bea
h_0^{mn}K_{mn}=h_0^{mn} K_{\0 mn}\;\; \mbox{or}\;\;  h_0^{mn}k_{ mn} =0
\la{Kcon}
\eea

For the full nonlinear analysis let us consider a $V^{mn}(h_{ab})$ where $V^{mn}$ is a yet-unknown function of $h_{ab}$ (to be determined shortly). 
Consider the Poisson bracket with the Hamiltonian:
\bea
[V^{mn}h_{mn}, H]_{PB}=\left(V^{ab}+\fr{\pa V^{mn}}{\pa h_{ab}}h_{mn}\right)\dot{h}_{ab}
\eea
Let us require $V^{mn}$ to satisfy
\bea
\left(V^{ab}+\fr{\pa V^{mn}}{\pa h_{ab}}h_{mn}\right)=h^{ab} \la{Veq}
\eea
One can solve this iteratively as follows. In the leading order, we choose 
\bea
V_\0^{mn}=h_0^{mn}
\eea
Substituting this into \rf{Veq} and comparing the zeroth and first order terms gives
\bea
V^{mn}=h_0^{mn}+\fr12 q^{mc}q_c^n+\cdots
\eea
Given that $[V^{mn}h_{mn},H]_{PB}= \mathscr{L}_{\pa_{x^3}}(V^{mn}h_{mn})$ one may gauge-fix
\bea
\dot{V}(=2nK)=2n_0 K_0
\eea
which amounts to the desired gauge-fixing:
\bea
K=K_0
\eea

\section{Renormalization of S-matrix}

With the preparation of the previous section, we are now ready to carry out two central 
steps to establish the renormalizability. The first is to show that the second fundamental form is reduced in the sense to be explained below. Whereas the mathematical duality between the Riemannian and totally geodesic foliation played a central role in the mathematical accounts of the reduction in \cite{Park:2014qoa,Park:2015qxa}, it did not really play a quantitative role
in the quantization in the flat background. This should be attributed to simplicity of the flat background. As we will see below, the duality will play an important role in the present case.  
The second task is to explicitly carry out the renormalization for a given amplitude. For this, we will revisit the one-loop two-point graviton amplitude considered in the flat background case; we will also comment on other amplitudes.

\subsection{reduction of $K_{mn}$}

The gauge-fixing of the shift vector $N_m=0$ makes the dynamics along the $x_3$ 
direction dependent only on the dynamics in the rest of the directions as one can see
through the $\dot{\pi}_{mn}$ field equation with $N_m=0,n=n_0$:
\bea
-\dot{\pi}^{mn}&=& n_0\sqrt{-h}\; (\cR^{mn}-\fr12 h^{mn}\cR)
+\fr12 n_0 \sqrt{-h}(\pi^{ab}\pi_{ab}-\fr12\pi^2 )h^{mn} \nn\\
&&\hspace{1in}-2n_0 \sqrt{-h}(\pi^{mb}\pi^n_{b}-\fr12\pi \pi^{mn} )
\eea
Although the fields have the bulk dependence, this aspect becomes no longer true for the physical states. Let us determine the physical states $|phys>$. We require them to be annihilated by the lapse function constraint\footnote{Inclusion of the total derivative terms does not affect the lapse constraint.}
\bea
\Big[-K^2+K_{mn}K^{mn}+\cR\Big]|phys>=0  \la{nconstrq}
\eea
The trace of the Einstein's field equation in the covariant variable $g_{\m\n}$ vanishes
\bea
R=\cR+K^2-K^{ab}K_{ab}+2\nabla_\a(n^\b \nabla_\b n^\a-n^\a \nabla_\b n^\b)=0 \la{Radmwbtq}
\eea 
Combining these two equations, one gets
\bea
\Big[\cR+\nabla_\a(n^\b \nabla_\b n^\a-n^\a \nabla_\b n^\b)\Big]|phys>=0,  \la{Rphy}
\eea
which will constrain the polarization tensor eventually, and
\bea
\Big[-K^2+K^{ab}K_{ab}  - \nabla_\a(n^\b \nabla_\b n^\a-n^\a \nabla_\b n^\b)\Big]|phys>=0 \la{Radmwbts}
\eea
One can check that these equations are satisfied by the Schwarzschild solution. We now require them to be satisfied as the physical state conditions even after including the fluctuation.
What we are up to is to expand \rf{Radmwbts} in terms of the fluctuation fields. Although \rf{Radmwbts} yields an infinite series, one may consider only the first non-vanishing order. 
For a reason to be explained shortly, we will consider \rf{Radmwbts} in the asymptotic regions of the radial direction. (This is in the perspective of treating the Schwarzschild geometry as perturbation to the flat geometry in the first quantization sense.) Then the first non-vanishing order is the quadratic order as we will show below.
 
Whereas the total derivative terms certainly contribute to the background part, they do not contribute to the fluctuation part of the constraint. The fact that they do not contribute to the fluctuation part can be seen as follows. 
Since $K=\nabla_\b n^\b$ and $K$ has been gauge-fixed to be $K=K_0$, the second term of $\nabla_\a(n^\b \nabla_\b n^\a-n^\a \nabla_\b n^\b)$ only contributes to the background part. The first term can be written
\bea
\nabla_\a (n^\b \nabla_\b n^\a)=\nabla_\a(c n^\a)=c\nabla_\a n^\a+n^\a\nabla_\a c 
\eea
where we have used the fact that $n^\a$ can be taken as the tangent vector of a geodesic due to the mathematical duality between the Riemannian foliation and totally geodesic foliation \cite{Park:2014qoa,Park:2015qxa}. (In other words, a manifold of Riemannian foliation - which our manifold is - with $\nabla_m n=0$ admits totally geodesic foliation. Therefore one may take  $n^\a$ as the tangent vector of the geodesic. The tangent vector of a geodesic, $u^\a$, satisfies $u^\b \nabla_\b u^\a=c u^\a$ with $c$ being a function of the parameter of the curve ($r$ for the present case).)   
The first term is again just a function of $r$; the second term is also a function of $r$.

As for the $K^{ab}K_{ab}$ part, one should expand it to the second order of the fields and consider the resulting expression at large $r$. Let us first contemplate the reason for considering the asymptotic region. The physical meaning of this is that one treats the Schwarzschild geometry as perturbation to the flat geometry in the first quantization sense; or one may say that we are considering renormalization associated with such asymptotic states.  (For example, the curved space propagator can be computed as a perturbative series of the flat space propagator. More detailed discussion can be found in Appendix C.) It is in the same spirit that the quantum field theoretic interactions are set aside when the physical state constraints are imposed (see the footnote on page 33 in ch. 15 of \cite{Weinberg2}). One can determine the full solution of the field equation by carrying out two perturbations at the same time: the first quantized and second quantized perturbations.

The expansion of $K_{mn}K^{mn}$ up to (and including) the second order is given by
\bea \la{Kmmsexp}
&&\hspace{1in} K_{mn}K^{mn}
= K_{0ab}K_{0cd}h_0^{ac}h_0^{bd}\nn\\
&&\hspace{.7in}-2K_{0ab}K_{0cd}h_0^{ac}q^{bd}+2K_{0ab}k_{cd}h_0^{ac}h_0^{bd}\\
&&+K_{0ab}K_{0cd}(2h_0^{ac}q^{be}q_e^{d}+q^{ac}q^{bd})-4K_{0ab}k_{cd}h_0^{ac}q^{bd}
 +h_0^{ac}h_0^{bd}k_{ab}k_{cd}\nn  
\eea
where $q^{mn}=q_{rs}h_0^{rm}h_0^{sn}$. The zeroth part in the first line is combined with other zeroth parts and vanishes. (In other words, the zeroth order terms cancel among themselves in \rf{Rphy} and \rf{Radmwbts}.) The linear part vanishes in the large-$r$ limit as one can check easily. The quadratic terms that contain the $q_{mn}$ factors similarly vanish in the limit. Finally the last term ${k_{ab}}{k_{cd}}h_0^{ac}h_0^{bd}$ survives in the limit, therefore 
the constraint leads to
\bea
{k_{mn}}=0 \la{kmn}
\eea

\subsection{renormalization: an example}

Let us revisit the one-loop two-point renormalization. Take the kinetic term without the trace piece: one may evaluate the path integral by taking the traceless propagator and gauge-fixing $h=0$.
\bea
\sqrt{-g}\cL_{kin} = -\fr12\sqrt{-g}\; {\N}_\g \f^{\a\b}{\N}^\g \f_{\a\b}
\eea
As in the flat case, we use a traceless propagator:
\bea
<\f_{\m\n}(x_1)\f_{\r\s}(x_2)>=P_{\m\n\r\s}\,\D(x_1-x_2) 
\eea
$P_{\m\n\r\s}$ denotes
\bea
P_{\m\n\r\s}=\fr12(g_{0\m\r}g_{0\n\s}+g_{0\m\s}g_{0\n\r}-\fr12 g_{0\m\n}g_{0\r\s})
\eea
and $\D$ satisfies
\bea
\nabla_0^\m \nabla_{0\m} \D(x_1-x_2)=\d^\4(x_1-x_2) \la{sl}
\eea
where $\nabla_{0\m}$ denotes the covariant derivative associated with $g_{0\m\n}$.
The perturbative analysis parallels the flat case in \cite{Park:2015ota} except in the following two technical complications. 
The first complication comes from $\D(x_1-x_2)$ which is defined above.
Although the explicit form of $\D$ is not needed for the divergence analysis, it will be needed for computation of the finite parts.\footnote{The analysis of the finite parts will be pursued elsewhere.} By invoking the usual background independence of the background field method, we can borrow the previous result in \cite{Park:2015ota} in which the one-loop counter-terms were obtained, after including the ghost contribution, as
\bea
\D \cL_{1loop} = -\fr12 \fr{\G(\e)}{(4\pi)^2}\sqrt{-g}\Big[ \fr{17}{15}R_{\m\n}R^{\m\n}-\fr{34}{15}R^2
\Big] \la{totalctr}
\eea
More precisely speaking, one needs to expand this around $g_{0\m\n}$ to the second order in the fluctuation fields in order to obtain the counter-terms for the two-point amplitude.

The other complication lies in the reduction of the 4D covariant quantities.
As in the flat case, the counter-terms will be expressed in terms of various 
contractions of $R_{\m\n\r\s}$. The reduction of these quantities deserves 
some attention. As commented in \cite{Park:2015ota}, this particular step is not necessary at one-loop. It will be needed, however, for higher-loop amplitudes such as a two-loop three-point amplitude.

The bulk quantities and the hypersurface quantities are related as follows (see, e.g., \cite{Aliev:2004ds}):
the components of the Riemann tensor are (the index $n$ is not to be confused with the lapse function $n$)
\bea
R_{mnpq}&=&R^\3_{mnpq}+K_{mq}K_{np}-K_{mp}K_{nq} \\
R_{3mnp}&=&N^l(R^\3_{lmnp}+K_{lp}K_{mn}-K_{ln}K_{mp})
           -n(\nabla_n K_{mp}-\nabla_p K_{mn})\nn\\
R_{m3p3}&=&N^l\Big[N^r(R^\3_{rmlp}+K_{rp}K_{ml}-K_{rl}K_{mp})
           -n(\nabla_l K_{mp}-\nabla_p K_{ml})\Big]  \nn\\
        &&  -n\Big(\mathscr{L}_{(\pa_{x^3}-N^q\pa_q)}K_{mp}+\nabla_m \nabla_p n\Big)            
            +n\Big(N^l(\nabla_m K_{lp}-\nabla_l K_{mp})+n K_{mr}K^r_{p}\Big);\nn
\eea
the components of the Ricci tensor and scalar, respectively, are
\bea
R_{mn}&=&\cR_{mn}-\fr1{n}\Big(\mathscr{L}_{(\pa_{x^3}-N^q\pa_q)}K_{mn}+\nabla_m \nabla_n n\Big)
         -KK_{mn}+2K_{ml}K^l_n  \nn\\
R_{m3}&=& N^l\Big[\cR_{ml}-\fr1{n}\Big(\mathscr{L}_{(\pa_{x^3}-N^q\pa_q)}K_{ml}+\nabla_m         
                 \nabla_l n\Big)-KK_{ml}+2K_{mr}K^r_n\Big]-n(\nabla_m K-\nabla_lK^l_m)\nn\\                
R_{33}&=&N^mN^n\Big[\cR_{mn}-\fr1{n}\Big(\mathscr{L}_{(\pa_{x^3}-N^q\pa_q)}K_{mn}
           +\nabla_m \nabla_n n\Big)-KK_{mn}+2K_{ml}K^l_n \Big]\nn\\
       &&  -n(\mathscr{L}_{\pa_{x^3}}K+\nabla_l\nabla^l n)  
       -n^2 K_{lr}K^{lr}+2n N^r \nabla^l\Big(K_{lr}-\fr12 h_{lr} K\Big)
\eea
and
\bea
R=\cR-K_{mn}K^{mn}-K^2-\fr2{n}\Big(\mathscr{L}_{(\pa_{x^3}-N^q\pa_q)}K+\nabla_m \nabla^m n\Big)
\eea
These expressions are before any gauge-fixing.
Let us illustrate the procedure by taking $R_{\k_1\k_2\k_3\k_4}$.
First, we enforce the gauge-fixings and constraints: 
\bea
N_m=0,n=n_0(r),\nabla_m n=0
\eea
The Riemann tensor expressions get substantially simplified:
\bea
R_{mnpq}&=&{\cal R}_{mnpq}+K_{mq}K_{np}-K_{mp}K_{nq} \nn\\
R_{3mnp}&=&-n_0(\nabla_n K_{mp}-\nabla_p K_{mn})\nn\\
R_{m3p3}&=&
          -n_0\mathscr{L}_{\pa_{r}}K_{mp}            
            +n_0^2 K_{mr}K^r_{p}   \la{rrt}
\eea
The result \rf{kmn} implies that one can replace $K_{mn}$ by
\bea
K_{mn}=K_{0mn}
\eea
Finally use the following relation
\bea
\cR_{mnrs}&=&\Big(\cR_{mr}-\fr14\cR h_{mr}\Big)h_{ns}
-\Big(\cR_{ms}-\fr14\cR h_{ms}\Big)h_{nr}\nn\\
&&+\Big(\cR_{ns}-\fr14 \cR h_{ns}\Big)h_{ms}
-\Big(\cR_{nr}-\fr14 \cR h_{nr}\Big)h_{ms}
\eea
It will be possible to absorb the counter-terms by the following metric redefinition almost as in the flat case, 
\bea
g_{\m\n}\ra g_{\m\n}+c_1  g_{\m\n}R+c_2 R_{\m\n}+c_3 s_{\m\n}  \la{gredef}
\eea
Because of the presence of the terms that contain $K_{mn}$-factors in \rf{rrt}, another tensor, $s_{\m\n}$, is present compared with the flat case. (We have more on this in the conclusion.) It will be  worthwhile to explicitly carry the procedure out for three-point amplitudes; however, this task will not be pursued here.

\section{Conclusion}

In this work we have extended the quantization of gravity in the flat background to a Schwarzschild black hole background.  
We have shown that the crucial gauge condition $K=0$ in the flat case is generalized to $K=K_\0$, a constraint arising from the gauge-fixing of the conformal part of the diffeomorphism. This constraint led to reduction of $K_{mn}$.  
The duality between the Riemannian and totally geodesic foliations and the interplay between the Lagrangian and Hamiltonian analyses were essential for the reduction. Renormalization of the one-loop two point amplitudes has been revisited; much of the flat case analysis could be carried over.

The fact that the reduction of the physical states occurs after gauge-fixing four gauge parameters is worth noting. In other words, regardless whether one adopts the approach of \cite{Park:2014tia,Park:2014noa} (in which the bulk and hypersurface gauge symmetry were gauge-fixed)  or that of \cite{Park:2015ota} (in which only the bulk gauge symmetry was fixed at the off-shell level), the reduction of the physical states takes place once the shift vector and the trace of the fluctuation metric are gauge-fixed.

There are several future directions. One direction is to carry out more complete
one-loop two-point renormalization including the finite terms and renormalization conditions. For this, one should fully evaluate the Feynman propagator and carry out the double perturbation theory mentioned in Appendix C. An elaborate analysis of scales and/or renormalization group flow would be involved.
One would presumably need to introduce 3D cosmological function terms (they will be used in the renormalization associated with the presence of $s_{\m\n}$ in \rf{gredef}) and we would expect the virtual boundary terms \cite{Park:2013vpa} to play a role.

Another direction is to analyze the running of the coupling constant, which would require computation of several three-point functions. The flat case has not been analyzed yet so one may consider the flat case first. The metric field redefinition may carry a deeper physical meaning; it would be interesting to explore this issue either in the flat case or the black hole case.

It would also be worthwhile to explore whether slight modifications of the present work could render it applicable to (A)dS black hole cases. A certain part of the modifications is straightforward and has been presented in Appendix B.
\\

We will report on some of these issues in the near future.

\vspace{.3in}

\ni {\bf Acknowledgments}

\ni 
I thank S. Sin for his hospitality during my visit to Hanyang University, Seoul. 
I greatly benefited from discussions with him.

\newpage
\appendix

\renewcommand{\theequation}{A.\arabic{equation}}
 \setcounter{equation}{0}
\section{Conventions and identities}

The signature is mostly plus: 
\bea
\eta_{\m\n}=(-,+,+,+)
\eea
All the Greek indices are four-dimensional
\bea
\a,\b,\g,...,\m,\n,\r...=0,1,2,3
\eea
and all the Latin indices are three-dimensional
\bea
a,b,c,...,m,n,r...=0,1,2
\eea
Our definitions of the Riemann tensor, Ricci tensor and Ricci scalar are
\bea
&&R^\r{}_{\s\m\n}\equiv \pa_\m \G^\r_{\n\s}-\pa_\n \G^\r_{\m\s}
                  +\G^\r_{\m\l}\G^\l_{\n\s}-\G^\r_{\n\l}\G^\l_{\m\s}\nn\\
&&R_{\m\n}\equiv R^\k{}_{\m\k\n}\quad,\quad    R\equiv R^\n_\n             
\eea
Although the Christoffel itself is not a tensor, its variation is and can be expressed in terms of the covariant derivative (see, e.g., \cite{Blau}): 
\bea
\d \G^\l_{\m\n}=\fr12 g^{\l\k}(\nabla_\m \d g_{\n\k}+\nabla_\n \d g_{\m\k}
          - \nabla_\k \d g_{\m\n})
\eea
One can also show that
\bea
\d R^\r{}_{\s\m\n}= \nabla_\m \d\G^\r_{\n\s}-\nabla_\n \d\G^\r_{\m\s}\quad,\quad
\d R_{\m\n}=\nabla_\r \d\G^\r_{\m\n}-\nabla_\n \d\G^\r_{\r\m}  \la{rtv}
\eea
The fluctuation metric $\f_{\m\n}$ is defined through
\bea
g_{\m\n}\equiv g_{0\m\n}+\f_{\m\n}
\eea
The indices of $\f_{\m\n}$ are raised and lowered by $g^{0\m\n},g_{0\m\n}$.
The following shorthand notations were used in some places:
\bea
\f\equiv g^{0\m\n}\f_{\m\n}\quad,\quad \f^\m\equiv \pa_\k\f^{\k\m}
\eea
For the refined application of the background field method, an additional shift 
$\f_{\m\n}\equiv \f_{{}_B\m\n}+h_{\m\n}$ was made:
\bea
g_{\m\n}=g_{0\m\n}+\f_{{}_B\m\n}+h_{\m\n}
\eea
The graviton propagator is given by\footnote{The ghost propagator is given by
\bea
<C_\m(x_1)\Cb_\n(x_2)>&=&g_{0\m\n}\,\D(x_1-x_2)
\eea
}
\bea 
<\f_{\m\n}(x_1)\f_{\r\s}(x_2)>&=& P_{\m\n\r\s}\, \D(x_1-x_2) 
\eea
where, for the traceless propagator,
\bea
P_{\m\n\r\s}\equiv \fr12\Big(g_{0\m\r}g_{0\n\s}+g_{0\m\s}g_{0\n\r}
              - \fr12g_{0\m\n}g_{0\r\s}\Big);
\eea
it satisfies 
\bea
P_{\m\n\k_1\k_2}P^{\k_1\k_2}{}_{\r\s}=P_{\m\n\r\s}
\eea
For a flat background,
\bea
\D(x_1-x_2)=\int \fr{d^4k}{(2\pi)^4}\fr{e^{ik\cdot (x_1-x_2)}}{i k^2}
\eea
As for the induced 3D metric:
\bea
 h_{mn}\equiv g_{\a\b}e^\a_m e^\b_n\quad,\quad 
\mathsf{h}^{\a\b}\equiv h^{mn}e^\a_m e^\b_n\quad,\quad 
g^{\a\b}=\ve u^\a u^\b+\mathsf{h}^{\a\b}
\eea
where $h^{mn}$ is the inverse of the induced metric 
\bea
e^\a_m\equiv \fr{\pa x^\a}{\pa y^m}
\eea
The 3D Ricci scalar, Ricci tensor and Riemann tensor are denoted respectively by
\bea
\cR,\cR_{mn},\cR_{mnpq}
\eea
The 3D fluctuation metric is introduce through
\bea
h_{mn}\equiv h_{0mn}+q_{mn}
\eea
The second fundamental form splits accordingly:
\bea
K_{mn}=K_{\0 mn}+k_{ mn}
\eea
where $K_{\0 mn}$ denotes the classical value and 
\bea
k_{mn}\equiv
\fr1{2n}\mathscr{L}_{\pa_{x^3}} q_{mn}
\la{K4defqqlinq}
\eea
after $N_m=0$ gauge.

\renewcommand{\theequation}{B.\arabic{equation}}
 \setcounter{equation}{0}
\section{AdS and dS black hole cases \la{SScase}}

Much of the analysis in section \ref{upn} can be carried over to the AdS and dS case
black holes.
\bea
S=\int d^4 x \sqrt{-g}\;(R-2\L) \la{EHq}
\eea
The $\g_{mn}$ field equation is (we have set $N_m=0$)
\bea
&& n(R^\3_{mn}-\fr12\g_{mn}R^\3)+\fr12 n \g_{mn}K^2 +\g_{mn}{\pa_{3}}K
  +\fr12  n\g_{mn}K_{rs}K^{rs} \nn\\
&&-2 nK_{mp}K^p{}_n - nK_{mn}K
-\g_{mp}\g_{nq}{\pa_{3}}K^{pq}=-\L n\g_{mn}
\label{gmneom}
\eea
Multiplying $h^{mn}$, one gets
\bea
-\fr12 n R^\3+\fr12nK^2-\fr12 n K_{rs}K^{rs}+3\pa_3 K-h_{pq}\pa_3 K^{pq}=-3n\L
\eea  
Combined with the $n$-field equation,
$R^{(3)}-K^2+K_{mn}K^{mn}=2\L$,
the equation above becomes
\bea
3\pa_3 K-h_{pq}\pa_3 K^{pq}+2n\L=2\pa_3 K+2n K_{pq}K^{pq}+2n\L=0
\eea

\renewcommand{\theequation}{C.\arabic{equation}}
 \setcounter{equation}{0}
\section{Schwarzschild geometry as perturbation to flat background}

In this appendix, we elaborate on the statement above \rf{Kmmsexp}, the rationale behind the consideration of the asymptotic $r$-region. To illustrate the idea and avoid inessential complications, let us consider a $\l \Phi^3$ scalar theory in the 
Schwarzschild background 
\bea
S=\int d^4x \sqrt{-g}\;\Big[- \fr12 g_0^{\m\n}\nabla_{0\m} \Phi \nabla_{0\n} \Phi-\fr{\l}{3!}\Phi^3  \Big]
\eea  
and its field equation,
\bea
g_0^{\m\n}\nabla_{0\m} \nabla_{0\n} \Phi-\fr12 \l \Phi^2=0 
\eea
where $\nabla_{0\m}$ denotes the covariant derivative whose connection is constructed out of $g_{0\m\n}$. The Laplacian is given by
\bea
\nabla^2 &=& -\fr1{f}\fr{\pa^2}{\pa t^2}
     +\Big(f\fr1{r} \fr{\pa^2}{\pa r^2}r+f' \fr{\pa}{\pa r} \Big)-\fr{\vec{L}^2}{r^2} \nn\\
     \vec{L}^2& =& -\left[\fr1{\sin\th}\fr{\pa}{\pa \th}\Big( \sin\th \fr{\pa}{\pa \th} \Big)
      +\fr1{\sin^2\th}\fr{\pa^2}{\pa \vf^2}\right]\nn\\
       \vec{L}^2Y_{lm}&=& l(l+1)Y_{lm}
\eea
where
\bea
f\equiv 1-\fr{2M}{r}
\eea
The solution of the field equation can be computed perturbatively by writing it as
\bea
\Phi(x)=\r(x)+\int_y \D_M(x-y)V_\Phi(y)\Phi(y)  \la{Phipert}
\eea
where the $\r$ field satisfies
\bea
\nabla_0^2\;\r(x)=0
\eea
and $\D_M(x-y)$ denotes the position space Feynman propagator satisfying
\bea
\nabla_0^2\D_M(x-y)=-\d^\4(x-y)
\eea
The field theoretic potential $V_{\Phi}$ is defined by
\bea
V_\Phi\equiv -\fr12 \l \Phi
\eea
It is possible to formulate the perturbation theory of scattering amplitudes in terms of \rf{Phipert} (see, e.g., \cite{Huang}). 
Treating the Schwarzschild geometry as perturbation to the flat geometry - which is a perturbation in terms of the black hole mass parameter $M$ - becomes relevant when one tries to compute $\D_M(x-y)$ perturbatively. 
More specifically, the curved space Laplacian can be split, after separating out the Minkowski piece from the rest, into two pieces:
\bea
\eta^{\m\n}\pa_\m\pa_\n+\Big[
\Big(-\fr{6M}{r^2}+\fr{4M^2}{r^3}\Big)\fr{\pa}{\pa r}
+\Big(-\fr{4M}{r}+\fr{4M^2}{r^2}\Big)\fr{\pa^2}{\pa r^2}
+\fr{2M}{r^3}\Lbf^2\Big] 
\eea
We can treat the term in the square bracket as a quantum mechanical perturbation
\bea
v_M\equiv  \Big(-\fr{6M}{r^2}+\fr{4M^2}{r^3}\Big)\fr{\pa}{\pa r}
+\Big(-\fr{4M}{r}+\fr{4M^2}{r^2}\Big)\fr{\pa^2}{\pa r^2}
+\fr{2M}{r^3}\Lbf^2
 \la{pot}
\eea
and compute $\D_M$ by employing quantum mechanical perturbation theory just as one computes $\Phi$ by employing the quantum field theoretical perturbation theory.
Overall, one computes the field $\Phi$ by employing a ``double perturbation theory": one in the mass parameter $M$ and the other in the quantum field theoretic interaction $V_\Phi$.

In the gravity case, one has an additional constraint, \rf{Radmwbts}, that does not exist in the scalar theory. The leading order of the quantum mechanical and field theoretic potentials leads to $k_{mn}=0$ as we have seen in the main body.

\newpage

\end{document}